\documentstyle[prl,aps,epsf,preprint]{revtex}

\def\be{\begin{equation}}
\def\ee{\end{equation}}

\def\sut{$SU_H(3)$~}
\def\vshift{\;\;\;\;\;}

\def\st{\scriptscriptstyle}

\begin{document}

\author{T.~K.~Kuo\thanks{tkkuo@physics.purdue.edu} , Sadek~W.~Mansour\thanks{mansour@physics.purdue.edu}, and Guo-Hong Wu\thanks{wu@physics.purdue.edu}}
\address{Department of Physics, Purdue University, West Lafayette, Indiana 47907}
\title{Triangular Textures for Quark Mass Matrices}
\date{}
\maketitle 
\begin{abstract}
The hierarchical quark masses and small mixing angles are shown to lead to a simple triangular form for the $U$- and $D$-type quark mass matrices. In the basis where one of the matrices is diagonal, each matrix element of the other is, to a good approximation, the product of a quark mass and a CKM matrix element. The physical content of a general mass matrix can be easily deciphered in its triangular form. This parameterization could serve as a useful starting point for model building. Examples of mass textures are analyzed using this method.
\end{abstract}

One of the most puzzling problems confronting the Standard Model (SM) concerns the family structure of quarks and leptons. To date there is very little understanding of the phenomena of hierarchical fermion masses and their mixing angles. A common approach to the problem is to postulate phenomenological mass matrices with certain simple textures. Usually these matrices are assumed to be hermitian~\cite{Fritzsch,RRR} for simplicity, although non-hermitian mass matrices have also been studied~\cite{Branco,Haussling}. In practice, non-hermitian mass matrices arise naturally in many models~\cite{nonHmodels}.

In this Letter, we suggest a new parameterization of fermion mass matrices which is non-hermitian and, more specifically, triangular in form. It should be emphasized that any mass matrix can be easily brought into a triangular form by a right-handed rotation, whereas it is non-trivial to make it hermitian.
Also, the condition of minimal parameterization~\cite{Haussling,Koide1,Falcone} can be satisfied for this type of texture. In fact, we believe that in the minimal parameter basis the triangular form is the simplest to study with all unphysical features eliminated from the start. 
If one considers the case when, say, the $U$-quark mass matrix is diagonal, but the $D$-quark mass matrix is triangular (which in general contains six real numbers and a phase), there will be only ten parameters in the two mass matrices. These account for the six quark masses, three mixing angles and a phase in the Cabibbo-Kobayashi-Maskawa (CKM) matrix. Furthermore, we will show that when the matrix is ``upper-triangular'', with zeros in the lower-left part, each of the matrix elements is approximately equal to the product of a quark mass and a CKM matrix element. Note also that one can always choose a basis where one type is diagonal and the other triangular.

Thus the triangular mass matrix offers not only the most economical, but also the most physical parameterization. Starting from any proposed mass matrix, a transformation into the triangular form enables one to read off the physical parameters immediately. Conversely, all viable mass matrices can be obtained there from by a suitable rotation. 
Before we proceed to study specific triangular textures, it is worth mentioning that there are four types of triangular textures with the three zeros in the upper-left, upper-right, lower-left, and lower-right corners of the mass matrix. All these can be transformed into each other by exchanging rows and columns, which amounts to a change of basis for the left-handed and right-handed quarks, respectively. The first two types do not yield simple relations in terms of the quark masses and mixing angles~\cite{Haussling}. The latter two are simply related by exchanging the first and third family right-handed quarks. We are naturally led to choose the texture with zeros in the lower-left corner of the mass matrix, as will become evident later. 

Let us now start by writing both the $U$- and $D$-type quark mass matrices in the upper-triangular form:
\be \label{eq:tri}
T_f=\left(\matrix{a_1&a_2e^{i \phi_1} &a_3 e^{i \phi_3}\cr 0&b_1&b_2 e^{i \phi_2}\cr 0&0&c}\right), 
\ee
where the phases are explicitly shown ($a_1$, $b_1$, and $c$ are chosen to be positive using phase rotations of the right-handed quarks). We have suppressed the index $f=U,D$ from the mass parameters in Eq.~(\ref{eq:tri}). As stated before, any mass matrix, $M$, can be rotated by a series of right-handed rotations into this upper-triangular form. For example, an appropriate rotation $U=R_{13} R_{23} R_{12}$ with $T=MU$, can make the (3,1), (3,2), and (2,1) elements vanish in succession. 

We turn to the diagonalization of the matrix $H_f=T_f T_f^{\dagger}=M_f M_f^{\dagger}$ which is given explicitly by
\be
{H_f}=\left(\matrix{a_1^2+a_2^2+a_3^2&
 a_2 b_1 e^{i\phi_1}+ a_3 b_2 e^{i(\phi_3-\phi_2)}& 
 c a_3 e^{i \phi_3}\cr  
a_2 b_1 e^{-i\phi_1}+ a_3 b_2 e^{i(\phi_2-\phi_3)}& 
b_1^2+b_2^2 & c b_2 e^{i\phi_2}\cr 
c a_3 e^{-i\phi_3} & c b_2 e^{-i\phi_2} & c^2}\right).
\ee
In the basis where $M_U$ is diagonal and $M_D$ is of the triangular form (\ref{eq:tri}), we can solve for $M_D$ in terms of quark masses, mixing angles and the $CP$-violating phase.
Recall that in this basis
$V_{\rm CKM}^\dagger H_D V_{\rm CKM} = D^2_D$, 
where $V_{\rm CKM}$ is the CKM matrix and 
$D^2_D\equiv {\rm diag}\left(m^2_d, m^2_s,m_b^2\right)$.
Denoting $x_{ij}\equiv \left(V_{\rm CKM} D^2_D V_{\rm CKM}^\dagger\right)_{ij}$, the following relations between the triangular mass matrix parameters and the physical parameters can be derived:
\begin{eqnarray} \label{rel}
c&=&\sqrt{x_{33}}= m_b V_{tb} \left(1+O(\lambda^8)\right),\nonumber\\
b_2 e^{i\phi_2}&=&x_{23}/c= m_b V_{cb}\left(1+O(\lambda^4)\right),\nonumber\\
a_3 e^{i\phi_3}&=&x_{13}/c= m_b V_{ub}\left(1+O(\lambda^4)\right),\nonumber\\
b_1&=&\sqrt{x_{22} - b_2^2}=m_s V_{cs}\left(1+O(\lambda^4)\right),\nonumber\\
a_2 e^{i\phi_1}&=&\left(x_{12}-a_3 b_2 e^{i(\phi_3-\phi_2)}\right)/b_1 =   m_s V_{us}\left(1+O(\lambda^4)\right),\nonumber\\
a_1&=&\sqrt{x_{11}-a_2^2-a_3^2}=m_dm_sm_b/(b_1c)
 =  \frac{m_d}{V_{ud}}\left(1+O(\lambda^4)\right),
\end{eqnarray}
where the last relation is derived by using the determinant of the mass matrix $a_1 b_1 c=m_d m_s m_b$, and $\lambda=0.22$ as in the Wolfenstein parameterization~\cite{Wolfenstein}. To be consistent with Eq.~(1), $V_{ud}$, $V_{cs}$, and $V_{tb}$ are all chosen to be positive.
In a nutshell, $M_D$ has the simple expression (to $O(\lambda^4)$),
\be \label{MD}
M_D\simeq \left(\matrix{m_d/V_{ud} &m_s V_{us}  
  &m_b V_{u b}\cr 
 0 & m_s V_{cs}  & m_b V_{c b} \cr 0 & 0 & m_b V_{tb}}\right),\hspace{0.2in} (M_U~{\rm diagonal}).
\ee
It is the hierarchical structure of the masses and the CKM matrix which entails this simple form for the mass matrix.
One can easily verify that $V^\dagger_{\rm CKM} M_D \approx {\rm diag}(m_d, m_s, m_b)$. 
Also, under phase transformations of the quark fields, $M_D$ is invariant in form if we make the replacement $m_d/V_{ud}\rightarrow m_d/V^*_{ud}$.
Given any mass matrix $M$, after a rotation into the upper-triangular form (\ref{eq:tri}), one can read off directly the approximate mass eigenvalues and mixings. Of course, exact solutions can also be obtained from Eq.~(\ref{rel}). We have compared numerical solutions with Eq.~(\ref{MD}) and found them to agree very well.

The relations~(\ref{rel}) can be easily inverted to give the physical quantities in terms of the triangular matrix parameters. Up to $O(\lambda^4)$ corrections:
$m_b\simeq c,
m_s\simeq b_1\left(1+\frac{1}{2}\lambda^2\right),
m_d\simeq a_1\left(1-\frac{1}{2}\lambda^2\right),
\left|V_{ub}\right|\simeq a_3/c,
\left|V_{cb}\right|\simeq b_2/c,
\left|V_{us}\right|\simeq a_2/b_1 \left(1-\frac{1}{2}{\lambda^2}\right).$
Also, we can compute the Jarlskog invariant~\cite{Jarlskog}
\begin{eqnarray}
J & = & \frac{-i}{2}\frac{{\rm det}\left[H_U,H_D\right]}{\prod{\left(m_i^2-m_j^2\right)}}
= \frac{a_2a_3b_1b_2c^2}{(m_b^2-m_s^2)(m_b^2-m_d^2)(m_s^2-m_d^2)}\sin\Phi_D\nonumber\\ 
& \simeq & \frac{a_2a_3b_2}{b_1c^2}\sin\Phi_D, 
\end{eqnarray}
where $\Phi_D=\phi_1+\phi_2-\phi_3$. CP violating effects depend only on the combination $\Phi_D$. In general, one can put the phase $\Phi_D$ in Eq.~(\ref{eq:tri}) at any of the (1,2), (2,2), (2,3), or (1,3) positions. Up to $O(\lambda^4)$ correction, $\Phi_D$ is equal to the $\gamma$ angle of the unitarity triangle, $\Phi_D\simeq \gamma \equiv {\rm arg}\left[-\frac{V_{ud} V_{ub}^*}{V_{cd} V_{cb}^*}\right].$

If we take $M_D$ to be diagonal while $M_U$ is assumed to have an upper-triangular form as in Eq.~(\ref{eq:tri}), the corresponding relations between the $M_U$ parameters and the physical parameters can be derived in the same way as before. 
The approximate form for $M_U$ (to $O(\lambda^4)$) is found to be 
\begin{eqnarray} \label{MU}
M_U&\simeq&\left(\matrix{m_u/V_{ud}&m_c V_{cd}^*&m_t V_{td}^*\cr0&m_c V_{cs}&m_t V_{ts}^*\cr0&0&m_t V_{tb}}\right),\hspace{0.2in} (M_D~{\rm diagonal}).
\end{eqnarray}
Again, here $V_{ud}$, $V_{cs}$, and $V_{tb}$ are chosen to be positive.
The Jarlskog invariant is now given by
\begin{eqnarray}
J&=&-\frac{a_2a_3b_1b_2c^2}{(m_t^2-m_c^2)(m_t^2-m_u^2)(m_c^2-m_u^2)}\sin\Phi_U
  \simeq -\frac{a_2a_3b_2}{b_1c^2}\sin\Phi_U,
\end{eqnarray}
where $\Phi_U=\phi_1+\phi_2-\phi_3$ and all variables refer to $M_U$. The CP phase $\Phi_U$ is related to the $\beta$ angle of the unitarity triangle:
$\Phi_U\simeq -\beta -\beta_s \equiv -\beta\left(1+O(\lambda^2)\right)$, where $\beta \equiv {\rm arg}\left[-\frac{V_{cd}V_{cb}^*}{V_{td} V_{tb}^*}\right]$ and $\beta_s={\rm arg}\left[-\frac{V_{ts} V_{tb}^*}{V_{cs} V_{cb}^*}  \right]$.

Note that for mass matrices with hierarchical eigenvalues and small mixings, it is always possible to rotate them into the upper-triangular form with the largest element at the (3,3) position. This is the main difference between triangular matrices with zeros in the lower-left and upper-right, the latter does not have this simple hierarchical structure~\cite{Haussling}.

If one starts with the case when neither $M_U$ nor $M_D$ is diagonal, one can convert both into hierarchical triangular forms through right-handed rotations. It may be first necessary to extract a large, common left-handed rotation from both $M_U$ and $M_D$, which cancels out in $V_{\rm CKM}$, to ensure that only the (3,3) element is large. This is illustrated in the first example below. Furthermore, the diagonal elements are simply the mass eigenvalues if $a_{1,2}\ll b_1$. The CKM matrix can then be obtained directly,
\be
V_{\rm CKM}=V_U^\dagger V_D,
\ee
where $V_U$ and $V_D$ are obtained from $M_U$ and $M_D$, as in Eq.~(\ref{MU}) and Eq.~(\ref{MD}). Both $V_U$ and $V_D$ are given approximately by $R_{23}\left(b_2 e^{i\phi_2}/c\right) R_{13}\left(a_3 e^{i\phi_3}/c\right) R_{12}\left(a_2 e^{i\phi_1}/b_1\right)$.

{\em Examples:\\}
We now show that triangularization is a very useful tool to study the physical content of any mass matrix. Three examples are given: the first is based on the democratic mass texture~\cite{Koide2}, the second deals with a realization of the Fritzsch texture. The last example is a new texture that we found based on \sut~horizontal symmetry.

\noindent
{\em - Democratic Mass Texture:}
The quark mass matrices are taken to be democratic (all the Yukawa couplings are the same for each quark sector). For example, the $U$-quark mass matrix given in~\cite{Fukugita} is
\be
M_U=\frac{K_U}{3}\left[\left(\matrix{1&1&1\cr 1&1&1\cr 1&1&1}\right)+\left(\matrix{-\epsilon&0&\delta\cr 0&\epsilon&\delta\cr -\delta&-\delta&0}\right)\right],
\ee
where the second term expresses small effects which violate the democratic texture ($\epsilon \ll \delta\ll 1$).
In the limit when $\delta,\epsilon\rightarrow 0$, this matrix (also $M_D$) is diagonalized by the unitary matrix
\be
A=\left(\matrix{\frac{1}{\sqrt{2}}&\frac{-1}{\sqrt{2}}&0\cr \frac{1}{\sqrt{6}}&\frac{1}{\sqrt{6}}&\frac{-2}{\sqrt{6}}\cr \frac{1}{\sqrt{3}}&\frac{1}{\sqrt{3}}&\frac{1}{\sqrt{3}} }\right).
\ee
We may then rotate $A\;M_U\;A^{-1}$ into the following triangular form:
\be
T_U\simeq \frac{K_U}{3}\left(\matrix{\frac{-\epsilon^2}{2 \delta^2}&-\frac{\epsilon}{\sqrt{3}}&-\sqrt{\frac{2}{3}}\epsilon\cr 0 & \frac{2}{3}\delta^2 & \sqrt{2} \delta \cr 0 & 0& 3}\right),
\ee
The mass eigenvalues ($m_u\simeq \frac{-\epsilon^2 K_U}{6 \delta^2}, m_c\simeq \frac{2}{9}\delta^2 K_U, m_t\simeq K_U$) can be easily read off from the diagonal elements and the $U$-quark mixing angles are simply given by $\theta_U^{12}\simeq -\frac{\sqrt{3}}{2}\frac{\epsilon}{\delta^2}$, $\theta_U^{23}\simeq \frac{\sqrt{2}}{3}\delta$, and $\theta_U^{13}\simeq -\frac{2}{3\sqrt{6}}\epsilon$, in agreement with the results in~\cite{Fukugita}. A similar analysis can be done for the $D$-quark mass matrix.

\noindent
{\em - Fritzsch Mass Texture:}
The quark mass matrices studied in~\cite{Froggatt} are given by
\be
M_{D,U}=\left(\matrix{0&\sqrt{m_1 m_2} e^{i\delta_1}&0\cr \sqrt{m_1 m_2} e^{-i\delta_1}&m_2&\sqrt{m_1 m_3}e^{i\delta_2}\cr 0&\sqrt{m_1 m_3}e^{-i\delta_2}&m_3}\right),
\ee
where $m_i (i=1,2,3)$ are the quark masses for the $U$ and $D$ sectors.
The corresponding triangular form after performing right-handed rotations is given by
\be
T_{D,U}\simeq\left(\matrix{-m_1\left(1+\frac{m_1}{2 m_2}\right)&\sqrt{m_1 m_2}\left(1-\frac{m_1}{2 m_2}\right) e^{i\delta_1}&m_1\sqrt{\frac{m_2}{m_3}}e^{i(\delta_1+\delta_2)}\cr 0 & m_2\left(1-\frac{m_1}{2 m_2}\right) & \sqrt{m_1 m_3}\left(1+\frac{m_2}{m_3}\right)e^{i\delta_2}\cr 0 & 0 & m_3}\right).
\ee
Note the appearance of $m_1$ at the (1,1) position. The CKM matrix which follows agrees with the results of~\cite{Froggatt}.

\noindent
{\em - New $SU(3)$ motivated Texture:}
Consider the following texture:
\be
{M_D}=m_b\left(\matrix{a&a&a e^{i \phi_d}\cr b&b+2a&2b+2a\cr 0&0&1}\right).
\ee
The $U$-type mass matrix is taken to be diagonal and real.
To examine whether this form of $M_D$ is viable, we first convert it into the triangular form:
\be \label{ourtri}
T_D = m_b\left(\matrix{\frac{2a^2}{B}& \frac{2a\left(a+b\right)}{B}& a e^{i\phi_d}\cr 0 & B & 2b+2a\cr 0& 0& 1}\right),
\ee
where $B=\sqrt{2b^2+4ab+4a^2}$.
Comparing Eq.~(\ref{ourtri}) to Eq.~(\ref{MD}), we obtain the following relations (up to $O(\lambda^4)$ corrections):
\begin{eqnarray}
a&\simeq&\sqrt{\frac{m_d m_s}{2 m_b^2}},\nonumber\\
b&\simeq&\frac{m_s}{\sqrt{2} m_b}\left(1-\sqrt{\frac{m_d}{m_s}}-\frac{m_d}{m_s}\right),\nonumber\\
\left|V_{cb}\right|&\simeq& 2(a+b) \simeq \sqrt{2}\frac{m_s}{m_b}\left(1-\frac{m_d}{m_s}\right),\nonumber\\
\left|V_{ub}\right|&\simeq& a \simeq \sqrt{\frac{m_d m_s}{2 m_b^2}},\nonumber\\
\left|V_{us}\right|&\simeq& \sqrt{\frac{m_d}{m_s}}\left(1-\frac{m_d}{2 m_s}\right).
\end{eqnarray}
Moreover, we have the prediction 
\be
\left|\frac{V_{us} V_{cb}}{V_{ub}}\right|\approx 2.
\ee

Numerically, using the central values of the $D$-quark masses at $M_Z$~\cite{RGE} and $\gamma\simeq 60^\circ$~\cite{Parodi}, we obtain 
\be
a_{\st D}=4.93\times 10^{-3},\vshift
b_{\st D}=1.60\times 10^{-2},\vshift
\phi_d=-60^\circ.
\ee
The CKM matrix is 
\be
\left|V_{\rm CKM}\right|=\left(\matrix{0.976&0.219&0.00493 \cr 0.219&0.975&0.0418\cr 0.00792&0.0414&0.999 \cr}\right),
\ee
in good agreement with the experimental values.
The Jarlskog parameter calculated from $M_D$ is given by $J=-4.4\times 10^{-5}\sin \phi_d$.

We close with a few concluding remarks. The known hierarchy of the quark masses and mixing angles leads to a very elegant triangular texture for the mass matrices. When either $M_U$ or $M_D$ is diagonal, one can directly read off the masses and the CKM mixing from the mass matrices. The one independent phase therein corresponds to the angles $\gamma$ and $\beta$ in the unitarity triangle, respectively. If both $M_U$ and $M_D$ are triangular, the CKM matrix is a simple product of two unitary matrices. This result is very useful for model building. Given any model mass matrix, triangularization offers an immediate criterion for its viability. We have illustrated this with several examples, where in each case the result was obtained quickly and simply. Clearly, one could also reverse the process by rotating the triangular form to generate viable model mass matrices. We hope that this parameterization can help suggest models which are both phenomenologically correct and theoretically justifiable. 
The application of triangular mass matrices to the charged lepton sector is immediate, whereas for the neutrino sector, the interpretation of the triangular mass texture may not be as simple due to the large mixings and the possibility of nearly degenerate neutrino masses. Work along this direction is in progress.

\acknowledgements

T.~K.~ and G.~W.~ are supported by the DOE, Grant no.~DE-FG02-91ER40681. S.~M.~ is supported by the Purdue Research Foundation.

\end{document}